\documentclass{ws-procs9x6}

\newcommand{\hess}{H.E.S.S.}
\newcommand{\mrm}{\mathrm}
\newcommand{\degr}{\hbox{$^\circ$}}
\newcommand{\arcmin}{\hbox{$^\prime$}}

\begin{document}
  
  \title{Observations of Galactic Gamma-Ray Sources with H.E.S.S.\ }
  
  \author{D. Berge for the H.E.S.S.\ Collaboration}

  \address{Max-Planck-Institut f\"ur Kernphysik\\
    P.O.~Box~103980\\
    D-69029~Heidelberg\\
    Germany\\
    berge@mpi-hd.mpg.de}
  
  \maketitle

  \abstracts{\hess\ results from the first three years of nominal
    operation are presented. Among the many exciting measurements that
    have been made, most gamma-ray sources are of Galactic origin. I
    will concentrate here on an overview of Galactic observations and
    summarise and discuss observations of selected objects of the
    different source types.}

  \section{Introduction}

  The High Energy Stereoscopic System (\hess) is a system of four
  imaging atmospheric Cherenkov telescopes which commenced full
  operation in the Khomas Highland of Namibia in December
  2003~\cite{HessStatusJim}, at an altitude of 1800~m. The experiment is
  run by an international collaboration of mostly European
  institutes. It is built for very-high-energy (VHE) gamma-ray
  astronomy, exploiting the energy range above 100~GeV up to several
  tens of TeV. Gamma rays are measured by means of Cherenkov light
  emitted in air showers of secondary particles that form whenever a
  primary gamma ray hits the earth's atmosphere and is being
  absorbed. Using Cherenkov images of air showers one can deduce the
  energy and direction of the primary particle.

  \begin{figure}[ht]
    \centerline{\epsfxsize=4.4in\epsfbox{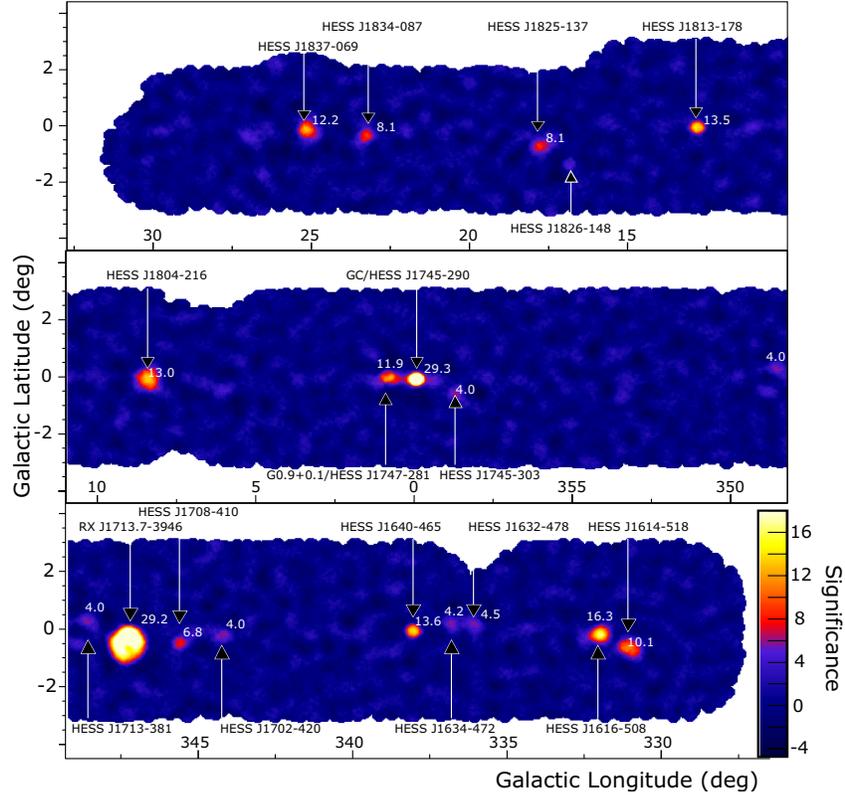}}
    \caption{Significance map of the \hess\ Galactic plane survey in
      2004~\protect\cite{ScanApJ}. The data include re-observations of
      gamma-ray candidates as well as pointed observations of known
      gamma-ray sources. The gamma-ray sources of the survey region
      are labelled and the significance of the signal is given for all
      of them. Note that the colour scale is truncated at
      $18~\sigma$.}
    \label{Scan_Image}
  \end{figure}

  The \hess\ Cherenkov telescopes are operated in moonless nights
  yielding a total observation time of roughly 1000~h per year. In
  normal data taking mode, five to ten objects are tracked per night
  with a typical cosmic-ray event rate of 300~Hz. The observations
  proceed in stereoscopic mode: events are recorded if at least two out
  of the four telescopes have triggered on the same air
  shower~\cite{TRIGGER}. The telescopes itself have a 60-t steel
  structure with \emph{altitude-azimuth} mount. Each has a tessellated
  mirror surface consisting of 380 single round facets, comprising a
  total area of $107~\mrm{m}^2$~\cite{KonradOptics}. With a focal length
  of 15~m, the Cherenkov light is imaged onto 960-photo-multiplier
  cameras with integrated fast readout
  electronics~\cite{PascalCamera}. Each camera covers a large field of
  view of $5\degr$. The resulting $FWHM \approx 4\degr$ of the system
  field-of-view response makes \hess\ the currently best suited
  experiment in the field for the study of extended VHE gamma-ray
  sources and the search for unknown sources in surveys.

  At zenith, the energy threshold of the system is about 100~GeV and for
  point sources an energy resolution of 15\% is achieved. The angular
  resolution for individual gamma rays is better than $0.1\degr$ and the
  point source sensitivity reaches $1\%$ of the flux of the Crab nebula
  for long exposures ($\approx 25$~hours).

  \section{The \hess\ Survey of the Inner Galaxy}
  One of the first observation campaigns of \hess\ in 2004 was a survey
  of the inner part of the Galaxy. Initially a total of 95~live~hours
  were recorded in scan mode, re-observations of promising gamma-ray
  source candidates yielded another 30~hours of data. Including pointed
  observations of the Galactic-centre region and the supernova remnant
  RX~J1713.7--3946 (which will both be discussed below), the \hess\ data
  set accumulates to 230~hours and reaches an average sensitivity of 2\%
  of the Crab flux above 200~GeV. In the region covered ($\pm 30\degr$
  in Galactic longitude, $\pm 3\degr$ in latitude) 14 previously unknown
  sources were detected. Fig.~\ref{Scan_Image} shows a map of the
  significance of gamma-ray emission of the survey region. 8 of the new
  sources exceed a significance level of $6~\sigma$
  post-trials~\cite{ScanScience}, 6 of them exceed the level of
  $4~\sigma$~\cite{ScanApJ}. They all line up with the Galactic plane,
  except for one all are extended at the 2 to $3\arcmin$ level and
  reveal hard power-law type energy spectra with a mean photon index of
  2.3.

  The \hess\ survey is a major breakthrough for the field of gamma-ray
  astronomy. The increased number of sources allows to consider the
  behaviour of population of sources, for the first time in this wave
  band. Using multi-wavelength observations one will now try to
  understand the physics of the acceleration processes that lead
  eventually to the emission of VHE gamma radiation. The sources in the
  survey region might be associated with four source classes:
  \begin{itemize}
  \item \textbf{Pulsar Wind Nebulae (PWNe):} HESS~J1825--137,\\
    HESS~J1747--281 (G0.9+0.1), HESS~J1702--420, and HESS~J1616--508.
  \item \textbf{X-ray binaries:} HESS~J1826--148 (LS~5039).
  \item \textbf{Supernova remnants (SNRs):} HESS~J1834--087,
    HESS~J1813--178, HESS~J1804--216, RX~J1713.7--3946, HESS~J1713--381,
    and HESS~J1640--465.
  \item \textbf{Unknown nature:} HESS~J1837--069, HESS~J1745-290
    (Galactic centre), HESS~J1745-303, HESS~J1708--410, HESS~J1634--472,
    HESS~J1632--478, HESS~J1614--518.
  \end{itemize}
  I will step now sequentially through the source classes and discuss
  examples of \hess\ measurements.

  \section{Pulsar Wind Nebulae}
  \begin{figure}[ht]
    \centerline{\epsfxsize=4.6in\epsfbox{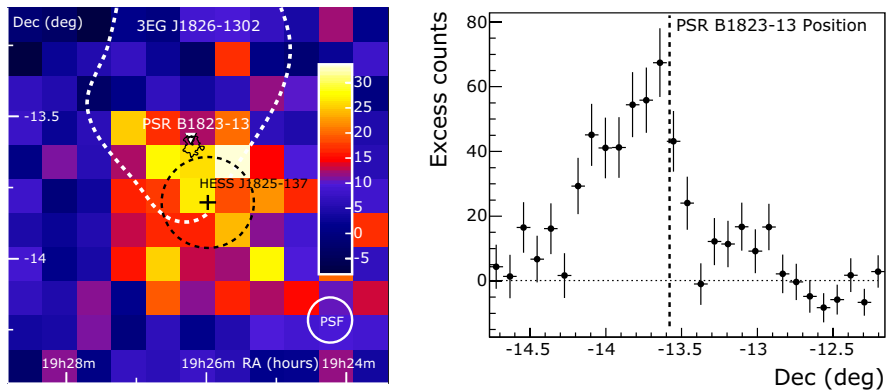}}
    \caption{\textbf{Left:} Gamma-ray excess image of the region
      surrounding PSR~B1823-13 (marked with triangle) in uncorrelated
      bins~\protect\cite{1825Paper}. The \hess\ best-fit position is
      shown with error bars together with the emission-region
      size. The black contours denote the XMM measurement, the dotted
      white line the unidentified EGRET source. \textbf{Right:} Excess
      slice ($0.4\degr$ wide) through the \hess\ data taken along the
      north-south direction. The one-sided nature of the emission with
      respect to the pulsar is clearly seen.}
    \label{J1825_Image}
  \end{figure}
  Energetic pulsars dissipate rotational energy in form of
  relativistic outflows. Confinement of these winds by the ambient
  medium leads to the formation of PWNe which can emit X-rays via
  Synchrotron radiation and gamma rays via the Inverse Compton
  mechanism. One of the four PWN candidates in the \hess\ survey
  region is HESS~J1825--137, shown in
  Fig.~\ref{J1825_Image}~\cite{1825Paper}. The source is probably
  associated with PSR~J1826--334, a $2.1 \times 10^4$ years old
  pulsar. As can be seen from the figure, the emission region is
  offset from the pulsar and extends asymmetrically to the south. The
  reason for this asymmetric PWN, which is also seen in the X-ray
  measurement, is the reverse shock from the northern side, where an
  increased density of the interstellar medium is encountered. The
  shock presumably crashed into the PWN and pushed it to the
  south. Note that follow-up observations of this object have been
  performed with \hess\ and more detailed analyses, including
  spatially resolved energy spectra, will be published very soon.

  \begin{figure}[ht]
    \centerline{\epsfxsize=4.6in\epsfbox{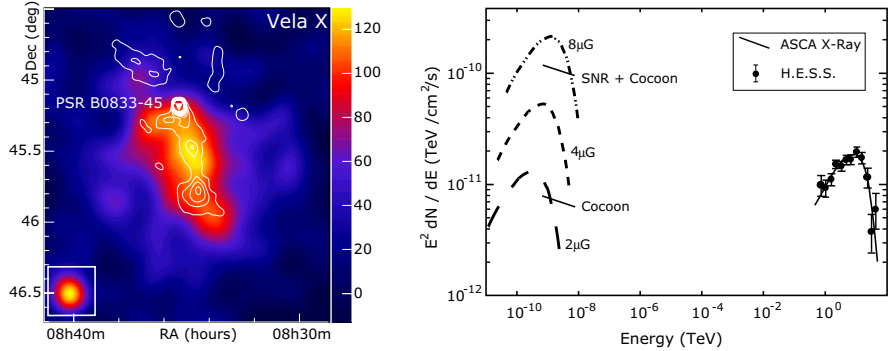}}
    \caption{\textbf{Left:} Gaussian smoothed gamma-ray image of the
      region surrounding the Vela pulsar~\protect\cite{VelaXPaper}
      (the pulsar position is marked with a triangle). The white
      contours are the ROSAT X-ray measurement of this region. In the
      bottom left-hand corner, a simulated point source is shown and
      demonstrates the resolution of \hess. \textbf{Right:} Spectral
      energy distribution using \hess\ and ASCA data. The black lines
      show one-zone model fits with different synchrotron flux
      predictions for different magnetic fields (see
      publication~\protect\cite{VelaXPaper} for details).}
    \label{VelaX_Image}
  \end{figure}
  Another example of a PWN measured in VHE gamma rays with \hess\
  (which is \emph{not} in the survey region) is Vela~X, the nebula
  associated with PSR~B0833--45. Fig.~\ref{VelaX_Image} shows the
  combined image from the 2004 and 2005 \hess\
  data~\cite{VelaXPaper}. The gamma-ray emission region is extended,
  roughly ellipsoidal in shape, and coincides well with ROSAT and
  Chandra X-ray measurements. Also here, the emission is offset from
  the pulsar, again due to an asymmetric reverse shock from the
  northern side. The energy spectrum measured with \hess\ is well
  explained by a one-zone Inverse Compton model, as is shown in
  Fig.~\ref{VelaX_Image}~(right). The measured photon index is very
  hard, $1.45 \pm 0.09$, with an exponential cutoff of $13.8 \pm
  2.3$. This is actually the first measurement of a complete VHE
  gamma-ray peak in a spectral energy distribution.

  \section{X-ray Binaries}
  \begin{figure}[ht]
    \centerline{\epsfxsize=4.6in\epsfbox{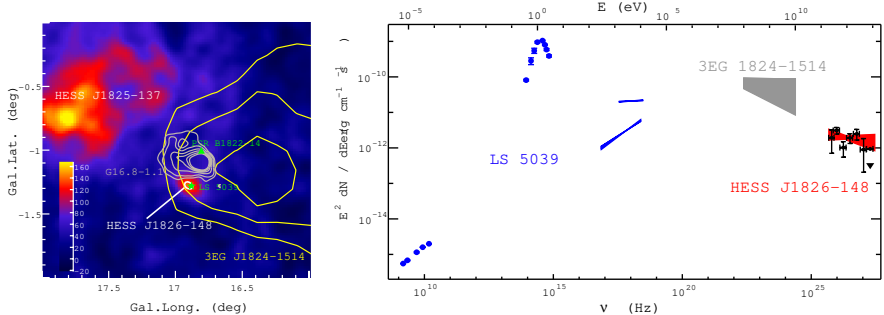}}
    \caption{\textbf{Left:} Smoothed excess image of the region around
      LS~5039~\protect\cite{5039Science}. The \hess\ position is
      indicated, overlaid are gray contours of radio emission and
      yellow contours of EGRET data. \textbf{Right:} Spectral energy
      distribution of LS~5039. \hess\ data (black points) are compared
      to optical and X-ray data. Shown in gray is the EGRET
      measurement suggesting an association of the \hess\ and the
      EGRET source.}
    \label{LS5039_Image}
  \end{figure}
  A point-like gamma-ray source was found close to HESS~J1825--137 in
  the \hess\ survey, HESS~J1826--148, likely to be associated with a
  system called LS~5039. This system is an X-ray binary, a companion
  star orbiting around a compact object. Radio and X-ray observations
  of relativistic outflows of some X-ray binaries have led to the term
  \emph{Microquasar}, suggesting that they behave as scaled-down
  active galactic nuclei. The \hess\ measurement is shown in
  Fig.~\ref{LS5039_Image}~\cite{5039Science}. It is noteworthy that
  this is the only point-like source in the whole survey region. The
  positional coincidence with LS~5039 led to the identification of the
  gamma-ray source with the microquasar, and it is the first detection
  of such an object in VHE gamma rays. The spectrum of HESS~J1826--148
  is shown in Fig.~\ref{LS5039_Image}~(right), it follows a power law
  and suggests an association with the EGRET source 3EG~1824--1514,
  despite a spatial separation of $\approx 0.5\degr$. More \hess\ data
  from follow-up observations in 2005 exist and allow to search for
  orbital modulations. Detailed results will be published soon.

  \section{Supernova Remnants}
  \begin{figure}[ht]
    \centerline{\epsfxsize=4.6in\epsfbox{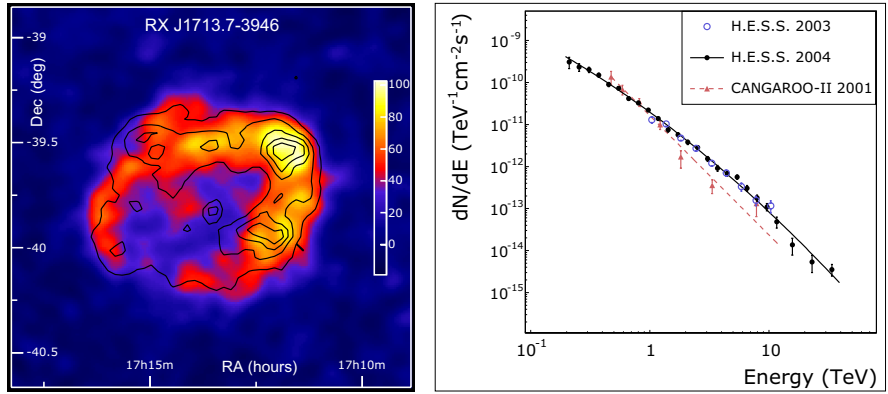}}
    \caption{\textbf{Left:} Smoothed gamma-ray excess image of
      RX~J1713.7--3946, produced from \hess\ data of 2004 and
      2005~\protect\cite{HESS1713c}. Note the angular resolution of
      $3.6\arcmin$ achieved here. Overlaid as black contours is the
      ASCA 1-3~keV X-ray measurement. \textbf{Right:} \hess\ gamma-ray
      spectrum of the whole SNR~\protect\cite{HESS1713b}. The black
      line is the best fit of a power law with photon index that
      depends logarithmically on energy, determined from the 2004 data
      set. The 2003 \hess\ data~\protect\cite{HESS1713a} shown as blue
      points are in good agreement. The CANGAROO-II data are also
      drawn.}
    \label{J1713_Image}
  \end{figure}
  SNRs are the best source candidates for cosmic rays in our
  Galaxy. The standard notion of particle acceleration is the
  diffusive shock acceleration of charged particles in the shells of
  SNRs. The source with the largest extension in the survey region is
  such a shell-type SNR, RX~J1713.7--3946. It has an apparent diameter
  of $\approx 1\degr$, twice the size of the full moon. The remnant
  was discovered with ROSAT in X-rays~\cite{ROSAT1713}, follow-up
  observations with ASCA revealed a dominantly non-thermal X-ray
  continuum without line emission~\cite{ASCA1713a,ASCA1713b}, most
  plausibly explained by Synchrotron emission of multi-TeV
  electrons. The presumed acceleration of electrons to TeV energies in
  the expanding shell of RX~J1713.7--3946, together with indications
  of interactions of the shock with molecular clouds~\cite{Fukui1713},
  made this SNR a prime target for \hess\ to look for gamma rays from
  interactions of accelerated cosmic rays with ambient matter.

  \begin{figure}[hb]
    \centerline{\epsfxsize=4.6in\epsfbox{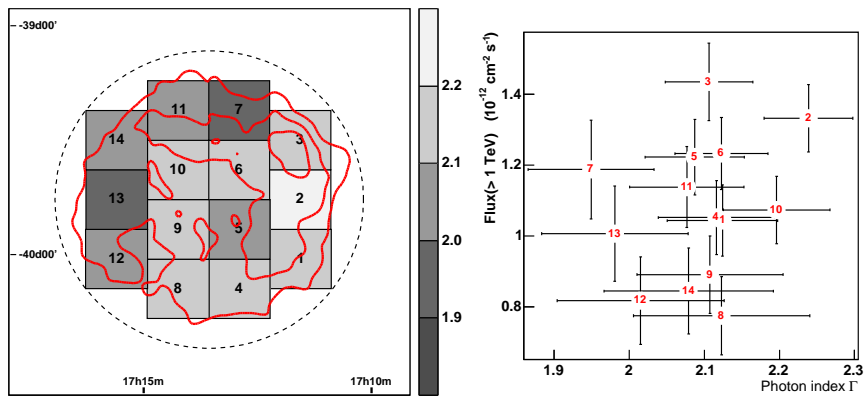}}
    \caption{\textbf{Left:} Gamma-ray excess contours are shown in
      red, superimposed are 14 boxes (each $0.26\degr \times
      0.26\degr$ in dimension) for which spectra were obtained
      independently~\protect\cite{HESS1713b}. The photon index
      obtained from a power-law fit in each region is colour coded in
      bins of 0.1. \textbf{Right:} Integral flux above 1~TeV versus
      the photon index, for the 14 regions shown left. The error bars
      are $\pm 1 \sigma$ statistical errors.}
    \label{J1713_IndicesRegions}
  \end{figure}
  After the first detection of VHE gamma rays from this object with
  CANGAROO~\cite{CANGI,CANGII}, \hess\ has indeed confirmed gamma-ray
  emission with its 2003 observation campaign. It revealed the first
  ever resolved image of an astronomical source in VHE gamma
  rays~\cite{HESS1713a}. Follow-up observations allowed for detailed
  analyses with unprecedented precision~\cite{HESS1713b}, the
  resulting gamma-ray image is shown in
  Fig.~\ref{J1713_Image}~\cite{HESS1713c}. It shows a clear shell
  structure, brighter in the northwest, resembling very much the
  picture seen in X-rays. In fact a detailed correlation study
  revealed a striking correspondence between keV and TeV energies. The
  differential energy spectrum of the whole remnant is shown in
  Fig.~\ref{J1713_Image}~(right). It extends over more than two
  decades well beyond 10~TeV and is well described by power-law type
  spectral shapes, albeit with deviations from a pure power law at
  large energies. The spectrum reported by the CANGAROO-II
  collaboration, also shown in the figure, is in marginal agreement
  with the \hess\ measurement.

  The \hess\ 2004 data of RX~J1713.7--3946 enabled us to perform a
  spatially resolved spectral analysis, to look for spectral variation
  on scales down to $\approx 0.3\degr$. The result is shown in
  Fig.~\ref{J1713_IndicesRegions}. When determining spectra in 14
  boxes arranged to cover the whole SNR, no significant index
  variation is found, the spectral shape is the same everywhere, only
  the flux varies by more than a factor of two.

\begin{figure}[hb]
    \centerline{\epsfxsize=4.6in\epsfbox{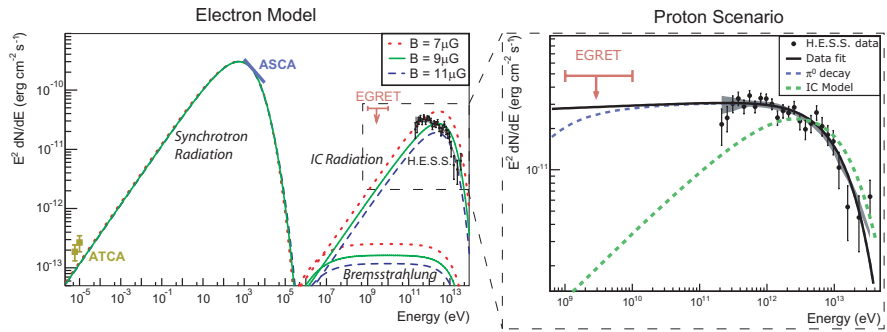}}
    \caption{\textbf{Left:} Spectral energy distribution of
      RX~J1713.7--3946. Shown are broadband data together with model
      curves obtained from a one-zone electron
      model~\protect\cite{HESS1713b}. Curves are plotted for three
      assumed magnetic field values. \textbf{Right:} Blow-up view of
      the high-energy part showing \hess\ data together with the fit
      of a power law with exponential cutoff, extrapolated to small
      energies. Moreover, a curve taking the gamma-ray suppression due
      to the $\pi^0$-decay kinematics into account is indicated and
      one of the Inverse Compton model curves from the left-hand
      side.}
    \label{J1713_ModelPlot}
  \end{figure}

  The key issue from the interpretation side for the RX~J1713.7--3946
  data is the identification of the particle population responsible for
  gamma-ray emission. While with the \hess\ measurement it is clear that
  primary particles are accelerated in the shock wave to energies beyond
  100~TeV, it remains difficult to say whether these particles are
  electrons or protons, in other words, if we really have the proof at
  hands that this SNR is a source of nucleonic cosmic rays. A broadband
  approach to answer this question is shown in
  Fig.\ref{J1713_ModelPlot}. A one-zone electron model fails to
  reproduce the spectral shape measured with \hess, in a hadronic
  scenario on the other hand the spectral shape seen in gamma rays is
  qualitatively as expected from theory. In that sense the hadronic
  scenario is favoured by the \hess\ data, although the correlation
  between X-rays and gamma rays is then challenging and so far not well
  understood.

  \begin{figure}[t]
    \centerline{\epsfxsize=4.6in\epsfbox{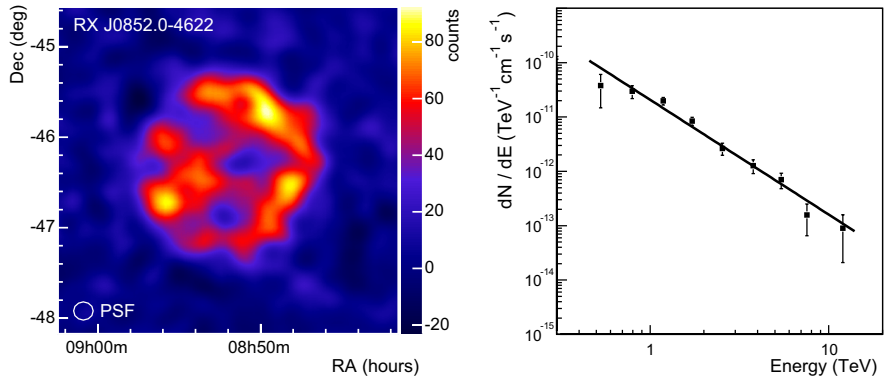}}
    \caption{\textbf{Left:} \hess\ gamma-ray excess image of
      RX~J0852.0--4622 from 2004 and 2005
      data~\protect\cite{VelaJrICRC}. The image is smoothed with a
      Gaussian of $\sigma = 0.1\degr$. The point-spread function (PSF)
      of this data set is shown in the bottom left
      corner. \textbf{Right:} \hess\ spectrum of the whole SNR from
      2004 data, determined from only 3.2~h live time (corresponding
      to $700 \pm 60$ excess events)~\protect\cite{VelaJrPaper}. The
      best-fit of a power law is shown as black line.}
    \label{VelaJr_Image}
  \end{figure}
  Another prominent SNR that was detected with \hess\ in 2004 is
  RX~J0852.0--4622~\cite{VelaJrPaper}, sometimes called \emph{Vela
  Junior} (it is close to the PWN Vela~X, discussed above). Also first
  discovered with ROSAT~\cite{ROSATVelaJr}, this object is in many
  regards similar to RX~J1713.7--3946. It is largely extended with a
  diameter of almost $2\degr$ and reveals a shell structure,
  correlated in X-rays and gamma rays. The \hess\ image is shown in
  Fig.~\ref{VelaJr_Image}. It demonstrates once more impressively the
  ability of \hess\ to map extended objects in gamma rays. The
  spectrum of the whole SNR is shown in
  Fig.~\ref{VelaJr_Image}~(right). It extends beyond 10~TeV and is
  within statistics well described by a pure power law with a photon
  index of $2.1 \pm 0.1$. Note that detailed analysis of more data
  from 2005 is underway and in the pipeline for publication.

  \section{Sources of Unknown Nature -- The \hess\ Galactic Centre Signal}
  \begin{figure}[ht]
    \centerline{\epsfxsize=4.6in\epsfbox{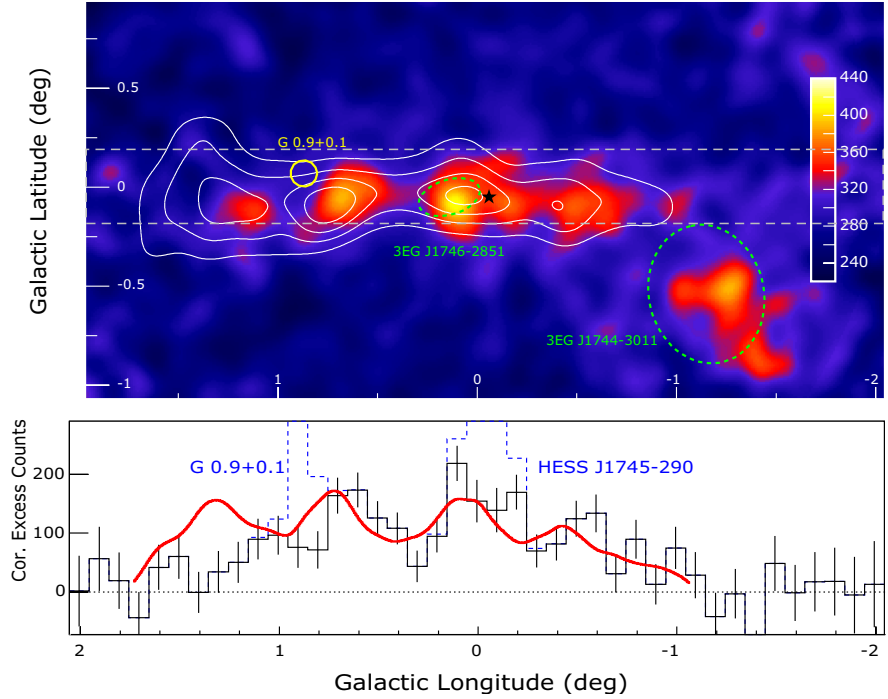}}
    \caption{\textbf{Upper panel:} Acceptance-corrected smoothed
      gamma-ray image of the Galactic centre region after subtracting
      the two dominant point sources in the field of
      view~\protect\cite{HESSDiffuse}. White contour lines indicate
      the density of molecular gas, traced by its CS emission. The
      dashed gray rectangle shows the $0.4\degr$ wide slice region
      that was used to produce the profile shown in the \textbf{lower
      panel:} Here we show the distribution of gamma-ray counts versus
      Galactic longitude and compare it to the CS line emission (red
      line). The signal of the two subtracted point sources is shown
      as dashed blue lines.}
    \label{GalCen_Image}
  \end{figure}
  Among all the \hess\ sources in the survey region that so far could
  not be unequivocally identified the Galactic centre is probably the
  most exciting one. The point-like VHE gamma-ray emission is
  coincident with the supermassive black hole $\mrm{Sgr~A}^*$ and the
  SNR Sgr~A~East~\cite{HESSSgrA}. The spectrum is well described by a
  pure power law with photon index $2.21 \pm 0.09$. No sign for any
  time variability of the signal is found. Possible emission processes
  that have been discussed include electron and proton origin of gamma
  rays, produced in the vicinity of the black hole or the shocks of
  the SNR. Moreover, the \hess\ signal has been discussed in the
  framework of dark matter annihilations~\cite{HESSDM}.

  \begin{figure}[th]
    \centerline{\epsfxsize=4in\epsfbox{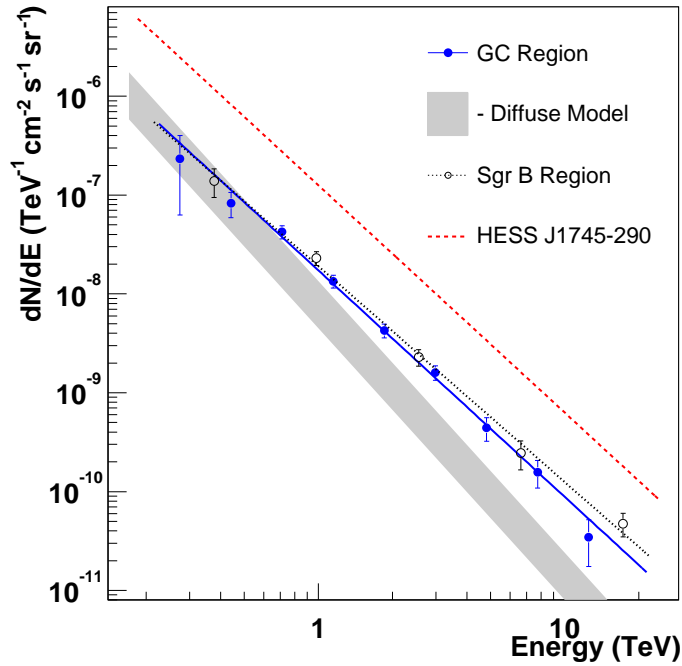}}
    \caption{Energy distribution of Galactic cosmic rays per unit
      angle in the Galactic centre
      region~\protect\cite{HESSDiffuse}. The spectrum is well
      described by a power-law fit (solid line). Data points are
      compared with the expected flux from $\pi^0$-decay assuming the
      local (solar) cosmic-ray spectrum and a target mass as measured
      with the CS emission. The open points correspond to the Sgr~B
      complex, the dotted red line gives the spectrum of the bright
      central source HESS~J1745-290.}
    \label{GalCen_Spectrum}
  \end{figure}
  The deep exposure of 2004 revealed not only a second source of VHE
  gamma rays, G0.9+0.1, but also enabled us to subtract these two
  strong point sources and search for remaining diffuse emission. The
  result is shown in the upper panel of Fig.~\ref{GalCen_Image} which
  shows the residual gamma-ray excess after subtraction. Two
  significant features appear: a region of extended emission spatially
  coincident with the unidentified EGRET source 3EG~J1744--3011, and
  emission extending along the Galactic plane for roughly
  $2\degr$~\cite{HESSDiffuse}. Overlaid in the figure are
  velocity-integrated CS data from the Galactic centre direction which
  trace molecular gas. There is a close correlation visible between
  the gamma-ray signal and the molecular gas density. In the lower
  panel of Fig.~\ref{GalCen_Image} the gamma-ray count rate is shown
  as a profile, plotted versus Galactic longitude, integrated in a
  $0.4\degr$ thick slice. The good match between gamma-ray and CS data
  suggests a cosmic-ray origin of gamma rays, produced in interactions
  of cosmic rays with molecular clouds. The similarity in the
  distributions of CS-line and gamma-ray emission implies a rather
  uniform cosmic-ray density in this
  region. Fig.~\ref{GalCen_Spectrum} shows the gamma-ray flux measured
  in this region. The data is well described by a power law with
  photon index $2.29 \pm 0.07$. Keeping in mind that in case of a
  power-law energy dependence, the gamma-ray spectral index closely
  traces the cosmic-ray index itself, it follows that the measured
  spectrum is significantly harder than in the solar neighbourhood. If
  we estimate the gamma-ray flux assuming a target mass as determined
  from the CS measurement, and the local cosmic-ray flux and spectrum,
  we obtain the shaded grey band shown in
  Fig.~\ref{GalCen_Spectrum}. There is a clear excess measured beyond
  500~GeV. This could simply be due to the proximity to the
  accelerator, meaning that propagation effects, which lead to a
  steepening of the spectrum, are less pronounced.

  \section{Summary and Conclusions}
  During its first three years of operation \hess\ has had a number of
  significant achievements in the field of VHE gamma-ray
  astronomy. Only with the sensitivity, the good angular and energy
  resolution and the large field of view of experiments like \hess\ is
  it now possible to measure the morphology and spectra of extended
  gamma-ray sources with great precision. Moreover, the good off-axis
  sensitivity make \hess\ ideally suited for sky surveys. This was
  demonstrated here by means of the Galactic plane survey data, which
  revealed 14 previously unknown VHE gamma-ray sources. The detection
  of extended emission from SNRs such as RX~J1713.7--3946, which
  resembles indeed as expected a shell structure, proves the existence
  of highest energy particles in the shocks of SNRs and presents a
  major step forward towards solving the puzzle of the origin of
  Galactic cosmic rays. Finally, the detection of a diffuse VHE
  gamma-ray component from the direction to the Galactic centre
  provides new vistas of the centre of our Galaxy delivering exciting
  insights into acceleration and diffusion processes of cosmic rays.

\end{document}